\documentclass[preprint,11pt]{elsarticle}

\usepackage{amssymb}
\usepackage[T1]{fontenc}
\usepackage{upquote}
\usepackage{xcolor}

\usepackage{color,soul}
\usepackage{listings}

\newcounter{bla}

\usepackage{hyperref}
\usepackage{algorithmic}
\usepackage{varwidth}
\journal{Journal of Computational Science}

\begin{document}
\definecolor{codegreen}{rgb}{0,0.6,0}
\definecolor{codegray}{rgb}{0.5,0.5,0.5}
\definecolor{codepurple}{rgb}{0.58,0,0.82}
\definecolor{backcolour}{rgb}{0.95,0.95,0.92}
\lstdefinestyle{mystyle}{
    backgroundcolor=\color{backcolour},
    commentstyle=\color{codegreen},
    keywordstyle=\color{magenta},
    numberstyle=\tiny\color{codegray},
    stringstyle=\color{codepurple},
    basicstyle=\footnotesize,
    breakatwhitespace=false,
    breaklines=true,
    captionpos=b,
    keepspaces=true,
    numbers=left,
    numbersep=5pt,
    showspaces=false,
    showstringspaces=false,
    showtabs=false,
    tabsize=2
}
\lstset{style=mystyle}
\begin{frontmatter}



\title{Performance evaluation of explicit finite difference algorithms with varying amounts of computational and memory intensity}


\author[a]{Satya P. Jammy}\corref{cor1}
\author[a]{Christian T. Jacobs}
\author[a]{Neil D. Sandham}

\cortext[cor1] {Corresponding author.\\\textit{E-mail address:} s.p.jammy@soton.ac.uk}
\address[a]{Aerodynamics and Flight Mechanics Group, Faculty of Engineering and the Environment, University of Southampton, University Road, Southampton, SO17 1BJ, United Kingdom}

\begin{abstract}
Future architectures designed to deliver exascale performance motivate the need for novel algorithmic changes in order to fully exploit their capabilities. In this paper, the performance of several numerical algorithms, characterised by varying degrees of memory and computational intensity, are evaluated in the context of finite difference methods for fluid dynamics problems. It is shown that, by storing some of the evaluated derivatives as single thread- or process-local variables in memory, or recomputing the derivatives on-the-fly, a speed-up of $\sim$2 can be obtained compared to traditional algorithms that store all derivatives in global arrays.
\end{abstract}


\end{frontmatter}

\section{Introduction}\label{sect:introduction}
Explicit finite difference methods are an important class of numerical methods for the solution of partial or ordinary differential equations. For example, they are used for numerically solving the governing equations  in computational fluid dynamics (CFD), astrophysics, seismic wave simulations, financial simulations, etc.

In CFD they are used by many researchers for the Direct Numerical Simulation (DNS) or Large Eddy Simulation (LES) of compressible flows. DNS is often performed to study boundary layers, aerofoils (involving both hydrodynamics and noise computations) \cite{Jones2012}, mixing analysis \cite{Pirozzoli2015}, shock-wave boundary layer interactions \cite{Wang2015} or benchmark test cases such as the Taylor-Green vortex \cite{DeBonis2013}, decaying homogeneous isotropic turbulence, etc. Even with the advances in computing hardware during the past decade, the current capabilities of DNS are limited to moderate Reynolds number flows \cite{Reynolds1990}.

It is expected that computing architectures will be capable of exaFLOPs ($10^{18}$ Floating Point Operations) by 2018 and 30 exaFLOPs by 2030 \cite{Slotnick2014}. Exascale architectures have the capability to perform DNS of the aforementioned examples (amongst others) at higher Reynolds numbers, or potentially wall-modelled LES  of the full model of an aircraft at operating Reynolds numbers. However, while there is a consensus \cite{Slotnick2014} that future architectures would not look much like the present IBM Blue Gene,  Cray, or IBM Productive, it is hard to predict the architectural design of such exascale systems. For example, they are expected to comprise less memory per core than the existing architectures. Exploiting the full potential of the exascale architectures poses many challenges to researchers, such as the sustainability of the solver's implementation with the uncertainty  of architectures, the need for new revolutionary algorithms/numerical methods, increasing computation  to communication ratio and the likelihood of I/O bottlenecks.

To address the problem of sustainability, taking into account the uncertainty in future architectures, one solution adopted by the CFD community involves  decoupling the work of a domain scientist and a computational/computer scientist \cite{JacobsPiggott2015}. In this approach, Domain Specific Languages (DSL) are developed by the computational/computer scientists, and the specifics of the problem and the numerical solution method are specified in the DSL by the domain scientist. Using source-to-source translation the numerical solver is targetted towards different parallel hardware backends (e.g. MPI, CUDA, OpenMP, OpenCL, and OpenACC) \cite{Rathgeber2012, Reguly2014}. This ensures that, for new architectures, only the backend that interfaces with the new architecture needs to be written and supported by the translator. The underlying implementation of the solver remains the same, thereby introducing a separation of concerns.

On the algorithms front, a lot of effort has gone into rewriting CFD solvers to exploit the available FLOPS of existing architectures. While the architectures have changed drastically in the last decade, algorithms have not advanced at a similar pace \cite{Slotnick2014}. Some algorithmical changes have been attempted by \cite{Salvadore2013, Thibault2009} to reduce the data transfer on Graphics Processing Units (GPUs), but a complete and detailed study on the performance of such algorithms on the existing CPU-based architectures is currently lacking. A first step towards exascale computing would be to evaluate the performance of algorithms characterised by varying intensities of memory usage and computational cost on current CPU-based architectures for a relevant hydrodynamic test case, solved using a finite difference scheme.

To facilitate these investigations, the capabilities of the recently developed OpenSBLI  framework \cite{JacobsSubmitted}  are extended to easily generate algorithms with varying amounts of computational and memory intensity.  OpenSBLI is a framework for the automated derivation and parallel execution of finite difference-based models. It is written in Python and uses SymPy to generate a symbolic representation of the governing equations and discretisation. The framework generates OPS-compliant C code that is targetted towards MPI via the OPS active library \cite{Reguly2014}.  A similar approach can also be applied to any set of compute-intensive equations solved using finite difference methods.

The aims of this paper are to: (a) study the performance of various algorithms on  current multi-core CPU-based architectures, (b) identify the best possible algorithm for the solution of explicit finite difference methods on  current multi-core CPU-based architectures, and (c) demonstrate the ease at which algorithmic manipulations can be achieved with OpenSBLI framework to overcome the challenges exascale architectures can pose.

The rest of the paper is organised as follows. The various algorithms are described in section \ref{sect:algorithms}. The validation of the algorithms is presented in section \ref{sect:validation}. The performance and scaling results are presented in section \ref{sect:performance}. Some conclusions are drawn in section \ref{sect:conclusion}.

\section{Algorithms}\label{sect:algorithms}
All the algorithms presented in this paper solve the three-dimensional unsteady compressible Navier-Stokes equations, with constant viscosity, given by
\begin{equation}
   \frac{\partial \rho}{\partial t} = - \frac{\partial}{\partial x_j}\left[\rho u_j\right],
\end{equation}
\begin{equation}
   \frac{\partial \rho u_i}{\partial t} = - \frac{\partial}{\partial x_j}\left[\rho u_i u_j + p\delta_{ij} - \tau_{ij}\right],
\end{equation}
and
\begin{equation}
   \frac{\partial \rho E}{\partial t} = - \frac{\partial}{\partial x_j}\left[\rho E u_j + u_j p - q_j - u_i\tau_{ij}\right],
\end{equation}
for the conservation of mass, momentum and energy, respectively. The quantity $\rho$ is the fluid density, $u_i$ is the velocity vector, $p$ is pressure and $E$ the total energy. The stress tensor $\tau_{ij}$ is defined as,
\begin{equation}
   \tau_{ij} = \frac{1}{\mathrm{Re}}\left(\frac{\partial u_i}{\partial x_j} + \frac{\partial u_j}{\partial x_i} - \frac{2}{3}\delta_{ij}\frac{\partial u_k}{\partial x_k}\right),
\end{equation}
where $\delta_{ij}$ is the Kronecker Delta and $\mathrm{Re}$ is the Reynolds number. The heat flux term $q_j$ is given by,
\begin{equation}
q_j = \frac{1}{(\gamma-1)\ \mathrm{M}^2\ \mathrm{Pr}\ \mathrm{Re}}\frac{\partial T}{\partial x_j},
\end{equation}
where, $T$ is temperature, $\mathrm{M}$ is the Mach number of the flow, $\mathrm{Pr}$ is Prandtl number and $\gamma$ is the ratio of specific heats. The pressure and temperature are given by,
\begin{equation}
\label{eq:pinprimitive}
p = \left(\gamma -1 \right)\left(\rho E - \frac{1}{2} \rho u_j^{2}\right),
\end{equation}
and
\begin{equation}
\label{eq:Tinprimitive}
T = \frac{\gamma \mathrm{M}^{2} p}{\rho},
\end{equation}
respectively. The variables that are advanced in time ($\rho, \rho u_i, \rho E$) are referred to as the conservative variables, and the right-hand sides (RHS) in the mass, momentum and energy equations are referred to as the residuals of the equations.

The mass, momentum and energy equations are discretised in space using a fourth-order central finite-difference scheme and a low storage Runge-Kutta (RK) scheme with three stages of temporal discretisation. For improved stability, the convective terms in the governing equations are rewritten using the formulation of \cite{Blaisdell1996},
\begin{equation}
\frac{\partial}{\partial x_j}\rho \phi u_{j} = \frac{1}{2} \left(\frac{\partial}{\partial x_j}\rho \phi u_{j} + u_{j} \frac{\partial}{\partial x_j}\rho \phi  + \rho \phi \frac{\partial}{\partial x_j}u_{j} \right),
\label{skewscalar}
\end{equation}
where $\phi$ is 1, $u_{i}$ or internal energy ($e$) for the mass, momentum and energy equations, respectively. To improve the stability of the present scheme, the viscous terms in the momentum and energy equations are expanded to second derivatives as used by \cite{Pirozzoli2015, Salvadore2013, Sandham2002}.

\begin{figure}
\noindent\fbox{%
\begin{varwidth}{\dimexpr\linewidth-2\fboxsep-2\fboxrule\relax}
\begin{algorithmic}
\STATE \texttt{set-the-initial-condition}
\FOR{\texttt{each-iteration}} \STATE {\texttt{save-state} }
\FOR{\texttt{each-rk-substep}}
\STATE {\texttt{evaluate-u\_i,p,T}}
\STATE {\texttt{evaluate-the-derivatives}}
\STATE {\texttt{evaluate-the-residual-of-the-equations}}
\STATE {\texttt{boundary-conditions}}
\STATE {\texttt{advance-solution-in-time}}
\ENDFOR \texttt{  //  end of rk sub loop}

\ENDFOR \texttt{  //  end of iteration loop}
\end{algorithmic}
\end{varwidth}
}
\label{fig:pseudocode}
\caption{Pseudo-code for the solution of the compressible Navier-Stokes equations.}
\end{figure}
A generic pseudo-code of the solution algorithm is shown in figure \ref{fig:pseudocode}. The time loop is the most computationally expensive part of the algorithm. It consists of evaluating the primitive variables ($p, u_i, T$), spatial derivatives, the residual for each equation and advancing the solution in time. This is achieved by iterating over the solution points of the grid, referred to as the grid loop in the rest of the paper. Various algorithms used for the evaluation of the residual of the equations are presented herein. Starting with a memory-intensive algorithm representing a typical handwritten CFD  solver, the amount of memory used and the computational intensity are varied, either by re-evaluating the derivatives on-the-fly or evaluating the derivatives using process-local variables. In all the algorithms presented, the primitive variables are evaluated and stored in memory.

\paragraph{Baseline algorithm (BL)}
This algorithm incorporates features similar to a typical handwritten static algorithm (i.e.~the derivatives in the residual of each equation are evaluated and stored in memory as arrays of grid point values; these are referred to as work arrays in the rest of the paper) on CPUs, to run as a sequential or parallel using MPI or OpenMP. For multi-threaded parallel programs, this requires the algorithm to be thread-safe in order to avoid race conditions; these occur when a variable is updated in the grid loop and the updated variable is used to update another variable in the same loop. For example, in the evaluation of the primitive variables from the conservative variables, the equation for pressure (\ref{eq:pinprimitive}) is dependent on the evaluated velocity components, and the equation for temperature (\ref{eq:Tinprimitive}) is dependant on the evaluated pressure. When running on threaded architectures, this potentially results in race conditions. This means that temperature could be evaluated before evaluating the pressure, and pressure could be evaluated before the velocity components are evaluated. Similar candidates for race conditions exist in the update equations (which advance the conservative variables forward in time) of the RK scheme.

To remove the race conditions, the code is generated such that no variable is updated and used in the same loop. This is achieved by separating the evaluations into multiple loops over grid points. For example, in the evaluation of primitive variables, the velocity components ($u_0, u_1, u_2$) are grouped into a single loop as the evaluations are independent, but the pressure and temperature are evaluated in different loops.

When generating the code that implements the BL algorithm, the first and second derivatives in the equations are evaluated and stored in work arrays in order to compute the RHS residual. The evaluation of the derivative of a combination of variables (e.g $\partial{\left(\rho u_{0} u_{0}\right)}/\partial{x_0}$) is achieved in two stages. In the first stage the function $\rho u_{0} u_{0}$ is evaluated and stored in a work array. In the second stage the derivative of the work array is evaluated using the central finite difference formula, and this result is stored in a new work array. The work array used in the first stage is not freed in memory, but is overwritten/reused  when evaluating other quantities.

The baseline algorithm is optimised such that computationally-expensive divisions are minimised. Rational numbers (e.g. finite difference stencil weights) and all the negative powers of constants in the equations are evaluated and stored at the start of the simulation. Typically, these are $\gamma^{-1}, \mathrm{Pr}^{-1}$, $\mathrm{Re}^{-1}$, and so on.
\lstset{style=mystyle}
\begin{lstinputlisting}[language=Python, caption=Key lines of the setup file for obtaining the BL algorithm., label=listing:BLA]
{tgv.py}
\end{lstinputlisting}

A sample setup file used to generate an implementation of this algorithm in OpenSBLI is shown in listing \ref{listing:BLA}. All the algorithms presented next are also optimised to reduce computationally expensive divisions. The setup file for other algorithms is similar to the BL algorithm with extra attributes to control the combinations of memory used and computational intensity.

\paragraph{Recompute All algorithm (RA)} In contrast to the BL algorithm, the evaluation of pressure and temperature are first rewritten using the conservative variables,
\begin{equation}
\label{eq:pinconser}
p = \left(\gamma -1 \right)\left(\rho E - \frac{1}{2} \rho \left(\frac{\rho u_j}{\rho}\right)^{2}\right),
\end{equation}
and
\begin{equation}
\label{eq:tinconser}
T = \frac{\gamma \mathrm{M}^{2} p}{\rho} =  \frac{\gamma\left(\gamma -1 \right) \mathrm{M} ^{2}\left(\rho E - \frac{1}{2} \rho \left(\frac{\rho u_j}{\rho}\right)^{2}\right)}{\rho} ,
\end{equation}
within the code to avoid race condition errors while fusing loops for the evaluation of the primitive variables.  Then, to evaluate the residual of the equation, all the continuous spatial derivatives in the residual are replaced by their respective finite difference formulas in the code generation stage. This differs from the BL algorithm in that, instead of evaluating the derivatives to work arrays and using them to compute the RHS residual, the code generation process directly replaces the derivatives by their respective finite difference formulas such that they are recomputed every time.

This algorithm results in a code in which no work arrays are used for storing the derivatives. The memory required for this algorithm is therefore the least of all algorithms and the computational intensity is the highest of all. The control parameters to generate code for this algorithm are shown in listing \ref{listing:RAA}.

\lstset{style=mystyle}
\begin{lstlisting}[language=Python, caption=Control parameters to generate the code for RA algorithm., label=listing:RAA]
grid = Grid(ndim,{"delta":deltas,"number_of_points":np})
grid.store_derivatives = False # Do not store derivatives
\end{lstlisting}

\paragraph{Store None algorithm (SN)}
This algorithm is similar to the RA algorithm. The difference is that, in the loop over grid points where the residuals are evaluated, each derivative in the RHS is evaluated to a single thread- or process-local variable. These variables are then used to update the residuals on a point-by-point basis, rather than storing all evaluations in a global-scope, grid-sized work array. To generate the code that implements this algorithm in OpenSBLI, the grid attribute \texttt{local\_variables} should be set to \texttt{True} along with the control parameters given in listing \ref{listing:RAA}. The pseudo-code for the residual evaluation as described here is provided in figure \ref{fig:pseudocode_SN}.

\begin{figure}
\noindent\fbox{%
\begin{varwidth}{\dimexpr\linewidth-2\fboxsep-2\fboxrule\relax}
\begin{algorithmic}
\FOR{\texttt{each-solution-point}}
\STATE {\texttt{double t1 = central difference formula for $\partial{u_0}/\partial{x_{0}}$}}
\STATE {\texttt{double t2 =  central difference formula for $\partial{u_1}/\partial{x_{1}}$}}
\STATE {\texttt{\vdots} }
\STATE {\texttt{residual = t1 + t2 \dots}}
\ENDFOR
\end{algorithmic}
\end{varwidth}
}
\label{fig:pseudocode_SN}
\caption{Pseudo-code for residual evaluation using SN algorithm.}
\end{figure}
The memory footprint of this algorithm is similar to that of the RA algorithm, but is slightly less computationally-intensive. This is because, for example, if a derivative is evaluated to a process-local variable then it can be reused if that derivative appears in any of the equations more than once.

\paragraph{Recompute Some algorithm (RS)}
In this algorithm, some of the derivatives (in this case, the first derivatives of the velocity components) are stored in work arrays and the remaining derivatives are replaced by their respective finite difference formulas in the residual. The evaluation of primitive variables follows the same procedure as the RA algorithm. Listing \ref{listing:RSA} shows the control parameters used to generate code for the RS algorithm. The memory usage for this algorithm is moderate, i.e. it is more than the RA algorithm but less than the BL algorithm.

\lstset{style=mystyle}
\begin{lstlisting}[language=Python, caption=Control parameters to generate the code for the RS algorithm, label=listing:RSA]
grid = Grid(ndim,{"delta":deltas,"number_of_points":np})
grid.store_derivatives = False # Do not store derivatives
grid.derivatives_to_store = set(problem.get_expanded_term_in_equations("Der(u_i,x_j)"))
\end{lstlisting}

\paragraph{Store Some algorithm (SS)} This algorithm is a fusion of the RS and SN algorithms, such that the derivatives that are not stored in the RS algorithm are evaluated and stored in thread- or process-local variables as per the SN algorithm. Listing \ref{listing:SSA} shows the control parameters used to generate code for this algorithm. Compared to the SN algorithm, an additional nine work arrays would be required for the SS algorithm for the 3D cases, and an additional four work arrays for 2D cases, since the first derivatives of the velocity components are now stored.

\begin{lstlisting}[language=Python, caption=Control parameters in setup file to generate the code for the SS algorithm., label=listing:SSA]
grid = Grid(ndim,{"delta":deltas,"number_of_points":np})
grid.store_derivatives = False
grid.derivatives_to_store = set(problem.get_expanded_term_in_equations("Der(u_i,x_j)"))
grid.local_variables = True
\end{lstlisting}

\section{Validation}\label{sect:validation}
The baseline (BL) algorithm is validated for a 3D compressible Taylor-Green vortex problem, to check the correctness of the solver. The initial conditions and  the post-processing procedure are described in \citep{DeBonis2013}. The simulations are performed in a cube of non-dimensional length $2 \pi$,  with periodic boundary conditions in all three directions for grids containing $64^3, 128^3, 256^3 $ and $512^3$ solution points. The Mach number, Prandtl number and Reynolds number of the flow are taken as $0.1$, $0.71$ and $1600$, respectively. The non-dimensional time-step for the $64^3$ grid size was set to $3.385 \times 10^{-3}$, and was halved for each increase in the grid size by a factor of $2^{3}$. Double-precision is used throughout all simulations presented in this paper.

\begin{figure}[!ht]
   \begin{center}
      \includegraphics[width=0.48\columnwidth]{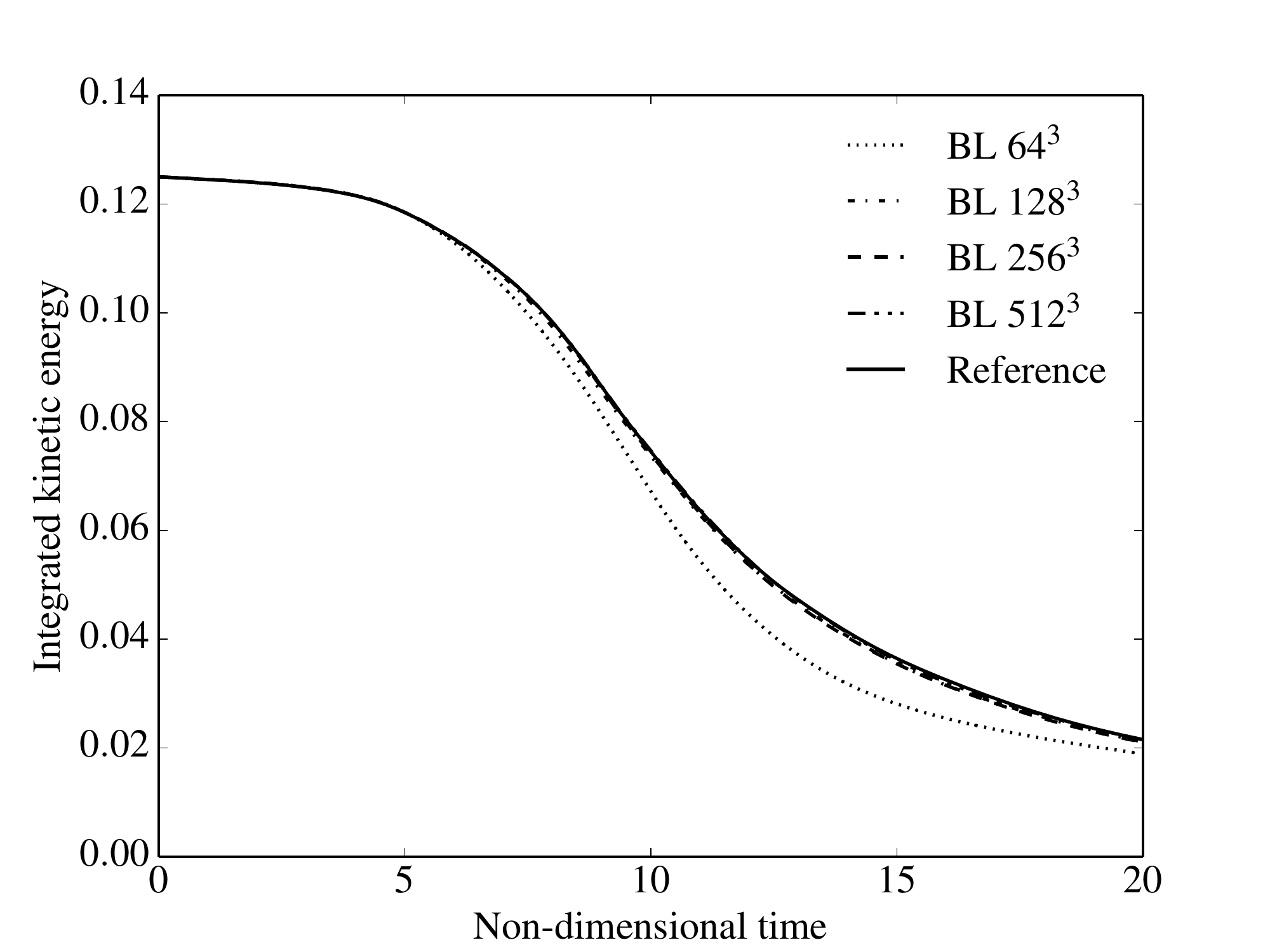}
      \includegraphics[width=0.48\columnwidth]{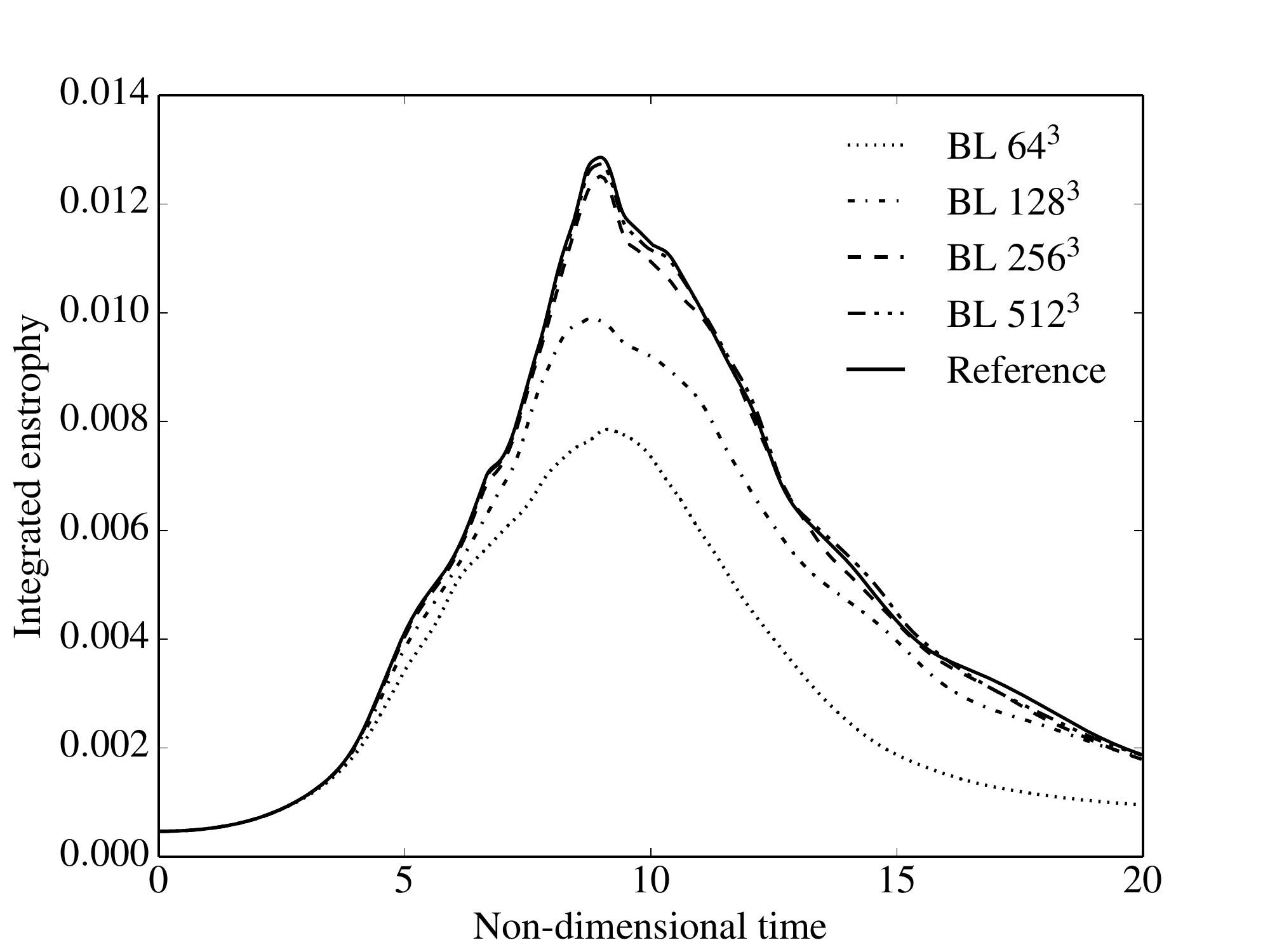}
      \caption{Left: Evolution of the integral of kinetic energy. Right: Evolution of the integral of enstrophy.}
      \label{fig:sf2ke_enstophy}
   \end{center}
\end{figure}

Figure \ref{fig:sf2ke_enstophy} shows the evolution of kinetic energy and enstrophy compared with the reference data \cite{Wang_etal_2013}. The results from the BL algorithm agree very well with the reference data for the $512^3$ case. For computational expedience, the other algorithms are validated on the $128^3$ grid. For each one, the results relative to the BL algorithm are found to be the same up to machine precision.

\section{Performance evaluation}\label{sect:performance}
After checking that the results from the various algorithms match, the performance of the different algorithms were evaluated using the same Taylor-Green vortex test case described in section \ref{sect:validation}. All simulations are performed on ARCHER (the UK National Supercomputing Service) and the code that implements the various algorithms is compiled using the Cray C compiler (version 2.4.2) with the -O3 optimisation flag. Each ARCHER node comprises 24 cores, with each MPI process being mapped to its own individual core. All simulations for performance evaluation purposes are run in parallel using 24 MPI processes/cores (one ARCHER node). The run-time of the time iteration loop was recorded for 500 iterations and is summarised in table \ref{Archerall} for a range of grid sizes.

\begin{table}[htbp!]
\centering
\begin{tabular}{|c|c|c|c|c|c|c|c|}
\hline
\textbf{Nx} & \textbf{Ny} & \textbf{Nz} & \textbf{BL} & \textbf{RA} & \textbf{RS} & \textbf{SN} & \textbf{SS} \\ \hline
64         & 64            & 64            & 16.21        & 9.29        & 10.76        & 8.44         & 9.78         \\ \hline
128        & 128           & 128           & 182.55       & 98.18       & 97.36        & 90.72        & 88.95        \\ \hline
256        & 256           & 256           & 1561.52      & 765.42      & 802.76       & 693.66       & 685.25       \\ \hline
\end{tabular}
\caption{Total run-time in seconds for different grid sizes for all algorithms on ARCHER using 24 MPI processes.}
\label{Archerall}
\end{table}
\begin{figure}[!ht]
   \begin{center}
      \includegraphics[width = 0.48\columnwidth]{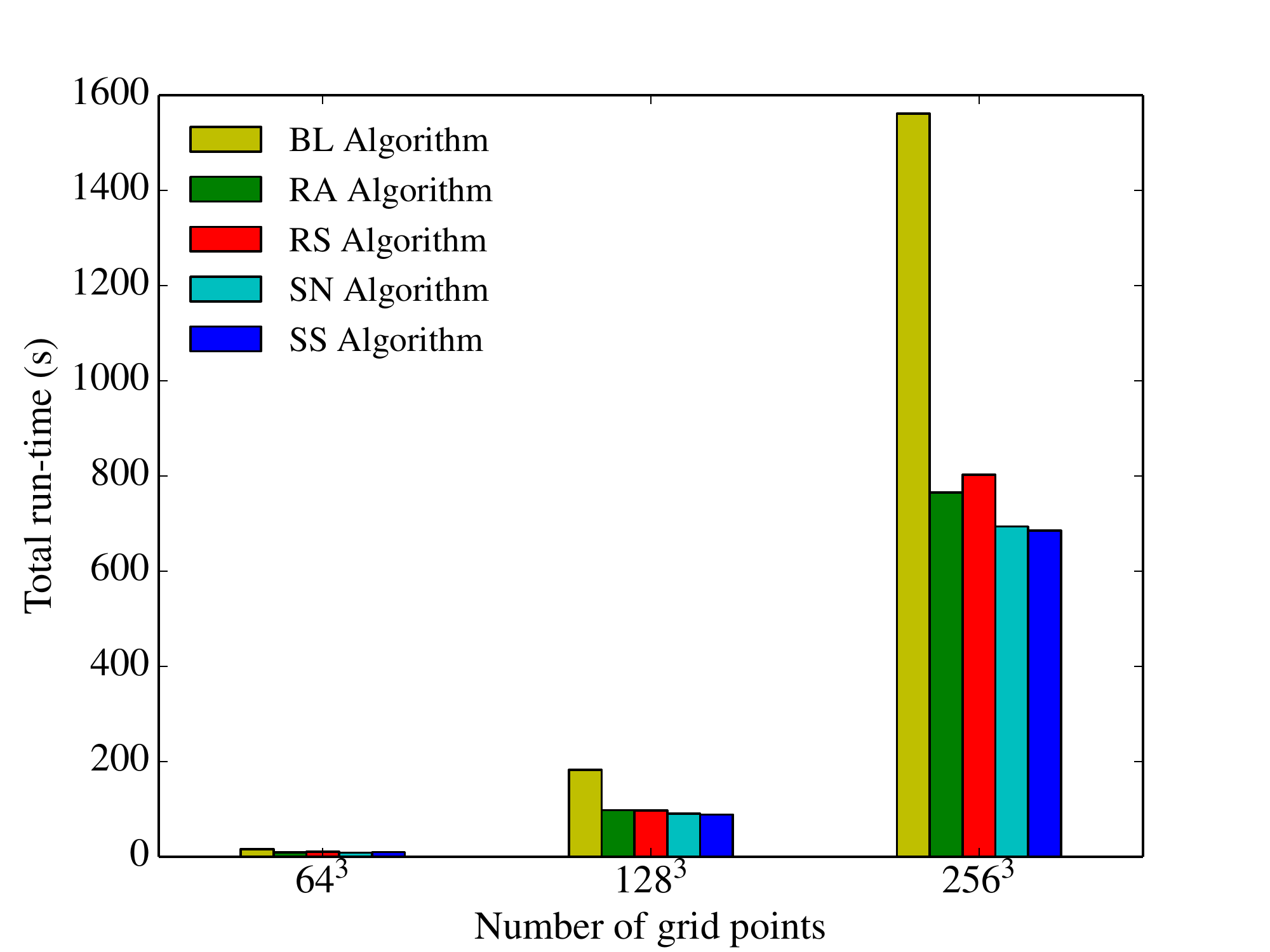}
      \includegraphics[width=0.48\columnwidth]{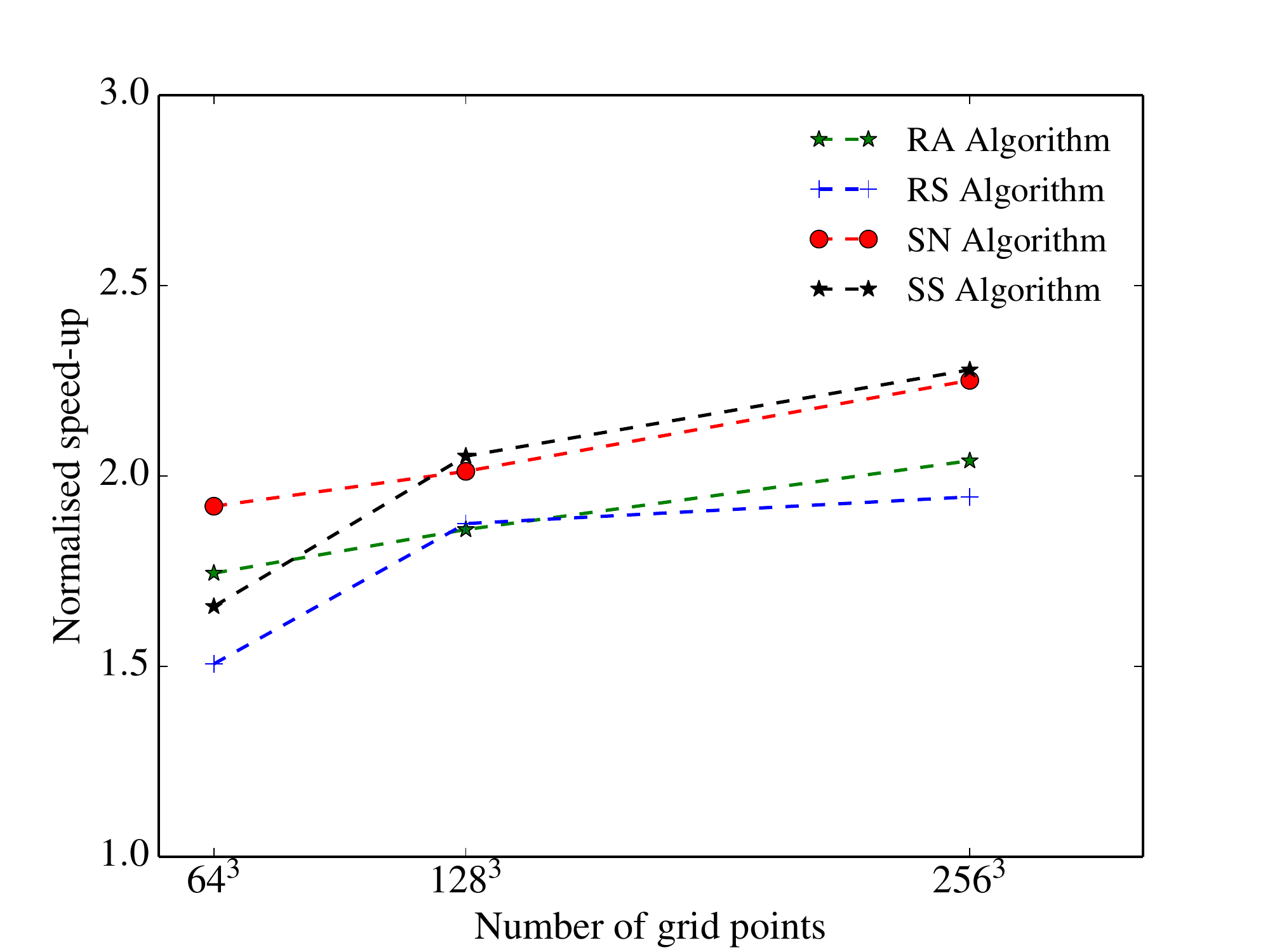}
      \caption{Left: Figure showing data in table \ref{Archerall}. Right: speed-up of algorithms normalised with the BL algorithm.}
      \label{fig:performance_archerall}
   \end{center}
\end{figure}
The data in table \ref{Archerall} is plotted in figure \ref{fig:performance_archerall}; from this figure it can be inferred that when the amount of memory access is reduced, the current CPU-based architectures perform better, even though the computational intensity of such algorithms is higher. The baseline algorithm is a factor of $\sim$2 slower than all the other algorithms presented in this paper. For larger grid sizes the benefit of the SS algorithm becomes more pronounced.

\subsection{Scaling}
\begin{figure}[!ht]
   \begin{center}
      \includegraphics[width = 0.8\textwidth]{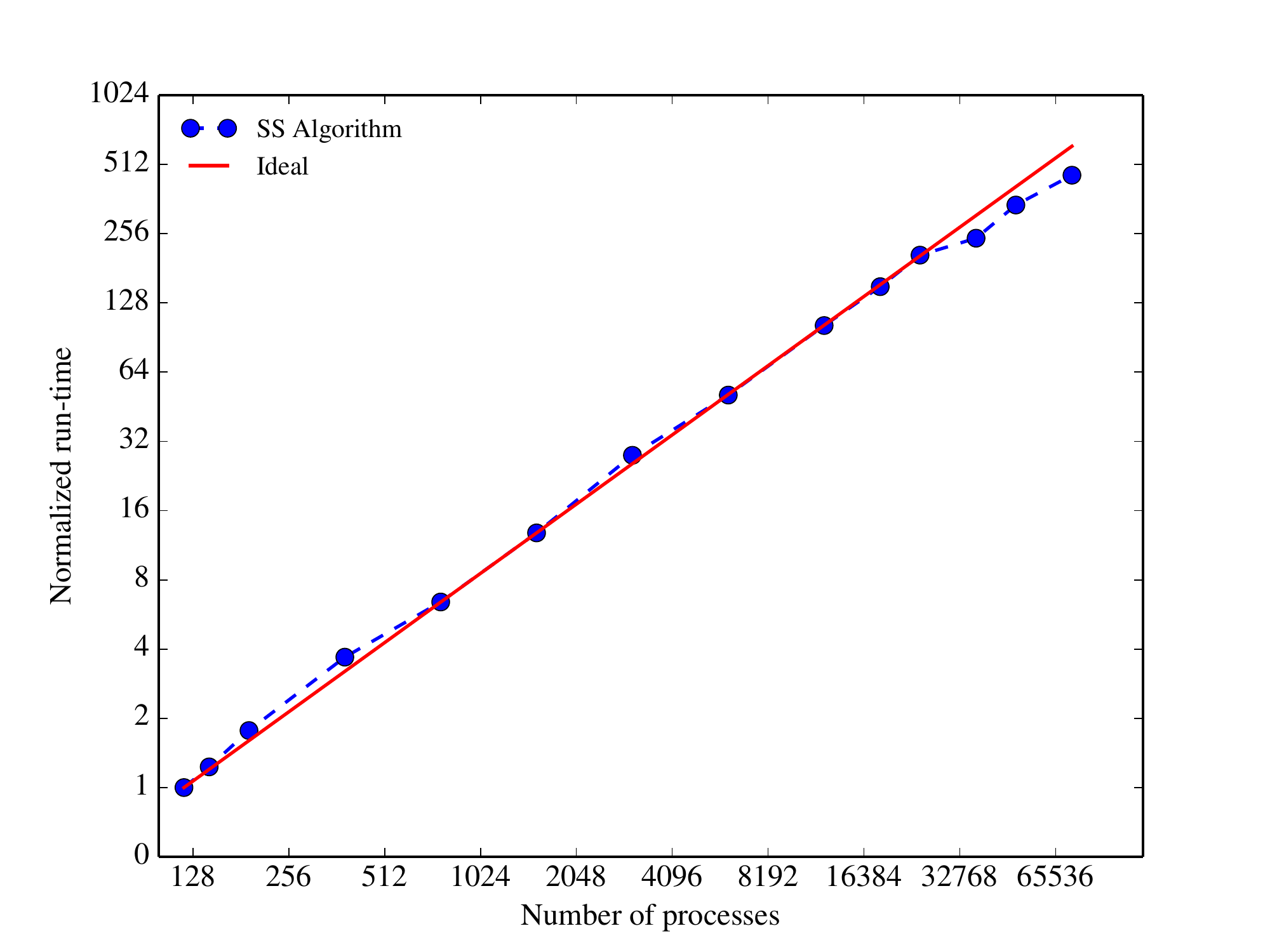}
      \caption{Strong scaling of the SS algorithm on ARCHER up to 73,728 cores using $1.07 \times 10^{9}$ grid points. The run-time has been normalised by that of the 120-process case.}
      \label{fig:OpenSBLI_strong_scaling}
   \end{center}
\end{figure}

Strong scaling tests were performed for the best performing algorithm (i.e. the SS algorithm) on ARCHER for the test problem with a total of $1.07 \times 10^{9}$ grid points and the runtime of the time iteration loop was recorded for 10 iterations. Figure \ref{fig:OpenSBLI_strong_scaling} shows the strong scaling results on ARCHER up to 73,728 MPI processes/cores (i.e. 3,072 ARCHER nodes). The minimum number of processes required for running the problem is 120. The algorithm shows a near-linear scaling (speed-up of 2) until 36,864 MPI processes (i.e. 1,536 ARCHER nodes) and thereafter the speed-up is $\sim 1.5$ as the process count doubles.

\begin{figure}[!ht]
   \begin{center}
      \includegraphics[width = 0.8\textwidth]{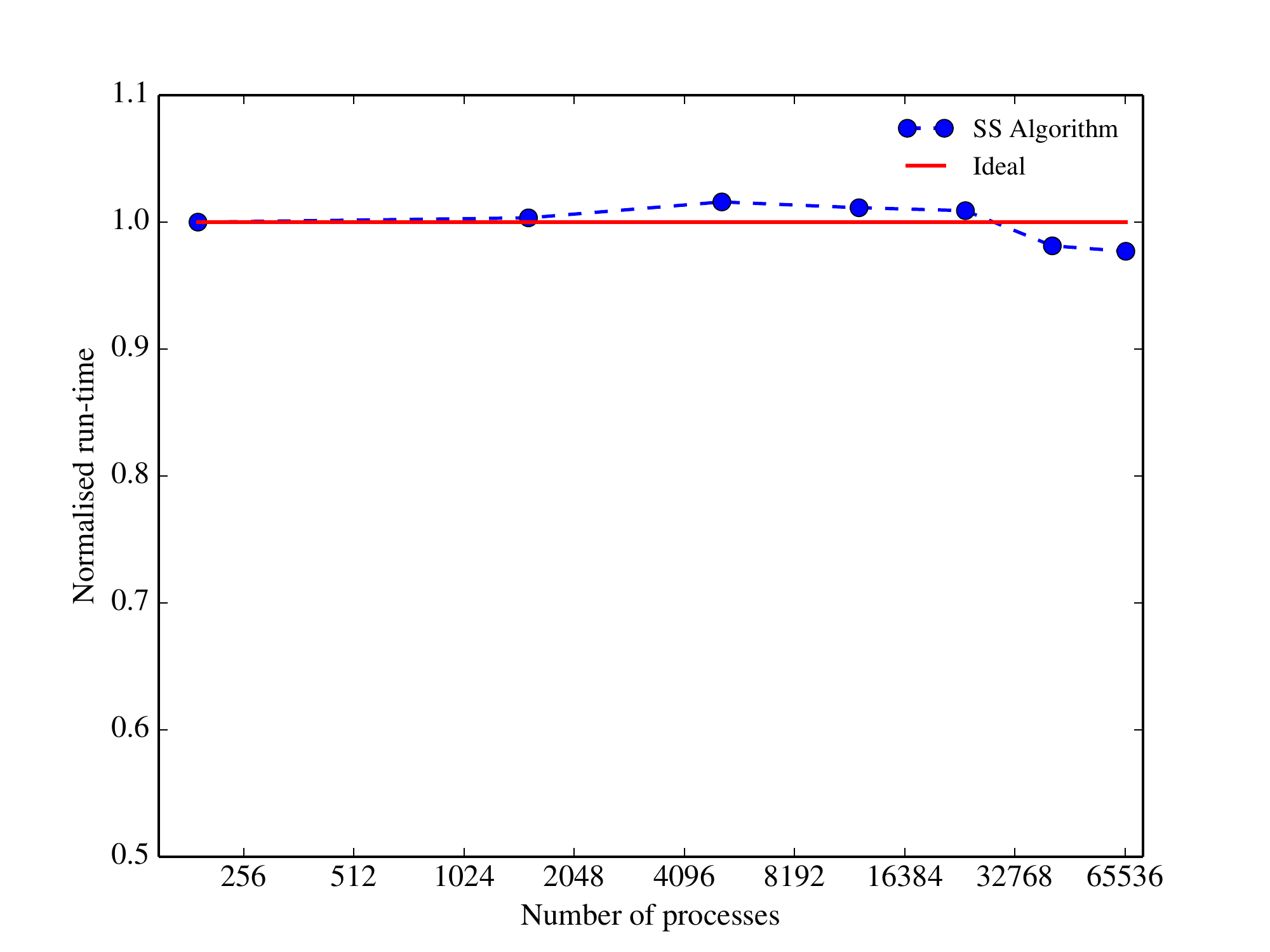}
      \caption{Weak scaling of the SS algorithm on ARCHER with $64^{3}$ grid points per MPI process up to 65,856. The results have been normalised by the run-time from the 192-process case.}
      \label{fig:OpenSBLI_weak_scaling}
   \end{center}
\end{figure}

Weak scaling tests were also performed for the SS algorithm. Here, the number of MPI processes was varied from 192 to 65,856 (i.e. 8 to 2,744 ARCHER nodes), while the number of grid points per MPI process was kept fixed at $64^3$ and the runtime of the time iteration loop was recorded for 10 iterations. The largest grid size considered comprises $\sim$17 billion solution points. Figure \ref{fig:OpenSBLI_weak_scaling} demonstrates that the normalised run-time is near-ideal.

\section{Conclusion}\label{sect:conclusion}
In this paper the automated code generation capabilities of the OpenSBLI framework have been extended to easily modify the memory usage and computational intensity of the solution algorithm. It was found that the baseline (BL) algorithm featured in traditional CFD codes, in which all derivatives are evaluated and stored in work arrays, is not the best algorithm in terms of performance on current multi-core CPU-based architectures. Recomputing all or some of the derivatives performs better than the baseline algorithm. The best algorithm found here for the solution of the compressible Navier-Stokes equations is to store only the first derivatives of velocity components in work arrays, and compute the remaining spatial derivatives and store them in thread- or process-local variables. The run-time of such an algorithm has been shown to be $\sim$2 times smaller than the BL algorithm. Through the use of modern code generation techniques in the OpenSBLI framework, it has been demonstrated that by changing just a few attributes (three in this case) in the problem setup file, different algorithms with varying degrees of memory and computational intensity can be readily generated automatically. The methodology presented in this paper can also be used to find the best possible algorithm for other existing architectures such as GPUs or Intel Xeon Phi coprocessors. Moreover, existing numerical models that use finite difference methods for the solution of any governing equations can be optimised for the current CPU-based architectures. When exascale systems become available, depending on their architecture and amount of available memory, users can readily tune the memory and computational intensity in the OpenSBLI framework to determine the best performing algorithm on such systems.

\section{Acknowledgements}
SPJ was supported by an EPSRC grant entitled ``Future-proof massively-parallel execution of multi-block applications'' (EP/K038567/1). CTJ was supported by a European Union Horizon 2020 project grant entitled ``ExaFLOW: Enabling Exascale Fluid Dynamics Simulations'' (grant reference 671571). The data behind the results presented in this paper will be available from the University of Southampton's institutional repository. The authors acknowledge the use of the UK National Supercomputing Service (ARCHER), with computing time provided by the UK Turbulence Consortium (EPSRC grant EP/L000261/1).



\section*{References}
\bibliographystyle{elsarticle-num}
\bibliography{opensbli}







\end{document}